\documentclass[runningheads]{llncs}

\usepackage[T1]{fontenc}

\usepackage{graphicx}

\usepackage[table]{xcolor}
\usepackage{booktabs}
\usepackage{tabularx}
\usepackage{threeparttable}
\usepackage{cite}
\usepackage{hyperref}

\emergencystretch=\maxdimen
\begin{document}

\title{ Functional Requirements for \\Decentralized and Self-Sovereign Identities }

\author{Daria Schumm\orcidID{0009-0004-1154-4799} \and
Burkhard Stiller\orcidID{0000-0002-7461-7463}}

\authorrunning{Schumm and Stiller}
\institute{ University of Zürich \\ 
Binzmühlestrasse 14, CH---8050 Zürich, Switzerland\\
\email{[schumm, stiller]@ifi.uzh.ch}}

\maketitle              

\begin{abstract}
    \vspace{-10px}
    Centralized identity management systems continuously experience security and privacy challenges, motivating the exploration of Decentralized Identity (DI) and Self-Sovereign Identity (SSI) as alternatives. 
    Despite privacy and security benefits to users, the adoption of DI/SSI systems remains limited. 
    One contributing reason is the lack of reproducible approaches to evaluate system compliance with its promised qualities. 
    Derivation of functional requirements (FR) is the first and necessary step to develop such an evaluation approach. 
    Previous literature on DI/SSI significantly lacks the systematic operationalization of existing non-functional requirements (NFR) or SSI principles. 
    This work addresses this research gap by deriving FR for a generalized DI/SSI use case, which encompasses the fundamental operations of the system. 
    The paper details operationalization methodology, introduces a formalized functional model, and presents a comprehensive set of FR, that can be used for future development and evaluation of DI/SSI systems. 
    As a result, establishing the fundamental step toward a reproducible evaluation framework, rooted in established requirements engineering methods. 

\end{abstract}


\section{Introduction}\label{sec:introduction}

    With the recent push toward mandatory governmental digital identity systems, public resistance has grown due to concerns about surveillance and data privacy \cite{a, b}. 
    These concerns continue to escalate with each new incident, such as the Optus and Ticketmaster data breaches \cite{d, 255}, exposing the risks of centralized identity management.  
    Decentralized Identity (DI) and Self-Sovereign Identity (SSI) offer an alternative by enabling data owners to retain control over their identity data without reliance on centralized services. 
    Despite the security and privacy benefits for users, adoption remains limited. 
    One contributing reason is the lack of transparent and reproducible approaches to evaluate system compliance with promised qualities (\textit{e.g.}, privacy or decentralization).
    Transparent evaluation and communication of compliance to users may facilitate greater trust in a new technology and support real-world adoption.

    To develop such an evaluation approach, concrete and measurable criteria for DI/SSI system functionality, or functional requirements (FR), must first be established. 
    Existing DI/SSI systems largely build on SSI principles, such as existence, control, and persistence, introduced by \cite{61}.
    These principles were extended and characterized as requirements that can be used as evaluation criteria by \cite{17}. 
    However, these properties represent non-functional requirements (NFR) and do not provide an operational or systematic basis for system evaluation. 
    Despite this limitation, most prior works (\textit{e.g.}, \cite{70, 85, 94, 245, 59, 72, 189, 79, 17, 191}) evaluate DI/SSI systems using NFR, which represent Cameroon's Laws of Identity \cite{198} and SSI principles \cite{61}.
    Other works (\textit{e.g.}, \cite{51, 87, 190, 44, 84, 32, 89, 6, 192, 69, 193, 102}) do not adhere to any of the requirements. 
    Across the literature, evaluation methodologies are often vague.
    Authors frequently neglect measurable metrics and fail to explain how the conclusions were derived, leading to ambiguous and non-reproducible results. 
    For example, \cite{245} and \cite{72} conclude that uPort lacks portability without clearly justifying how this conclusion has been reached. 
    Similarly, \cite{189} argues that the requirement of “consistent experience across context” for Sovrin is “hard to say” due to future design choices. 
    Even when explicit evaluation frameworks are proposed, such as in \cite{59}, systems are still assessed using NFR without systematic mapping to functionality, leaving substantial room for interpretation of the outcomes. 

    While NFR are valuable for expressing desired system qualities, relying on them as a primary evaluation criterion leads to unclear results, poor reproducibility, and limited transparency. 
    In contrast, FR define system functionality by describing the relationship between input, output, and state, and by specifying what a system should do to satisfy a certain NFR.
    Yet only a few works, namely \cite{186} and \cite{249}, specify FR in the DI/SSI domain. 
    \cite{186} identifies a small set of FR, derives use cases, and evaluates existing DI/SSI solutions, but does not provide mapping between FR and NFR, nor do the authors describe evaluation methodology. 
    \cite{249} outlines a scenario-based functional evaluation approach, but limits the scope to capability, compatibility, and interoperability, omitting key DI/SSI qualities, such as autonomy and consent. 
    As a result, retaining uncertainty about whether the evaluated functionality adequately reflects SSI principles.
    
    Overall, existing DI/SSI evaluations primarily rely on NFR or abstract system qualities. 
    While some functional representations of NFR have been specified in prior work (\textit{e.g.}, \cite{186, 249}), NFR are rarely operationalized systematically.  
    Consequently, existing evaluation methods (\textit{e.g.}, \cite{245, 59, 189}) yield ambiguous results, lack reproducibility, and fail to associate observed system behavior with the system quality it represents. 
    To address this research gap, this work derives FR, building on NFR derived from existing literature (\cite{17}; outlined in Table \ref{tab:nfr}).
    Derived FR represent a generalized DI/SSI use case, encompassing the fundamental operations of the system. 
    As a result, this work presents the first step toward building a transparent and reproducible evaluation framework.


    \begin{table}[]
    \centering
    \caption{Non-Functional Requirements of DI/SSI Systems}
    \label{tab:nfr}
    \def\arraystretch{1.5}%
    \footnotesize
    \rowcolors{2}{white}{gray!10}
    \begin{tabularx}{\textwidth}
        {>{\hsize=0.1\hsize\linewidth=\hsize}X 
         >{\hsize=0.2\hsize\linewidth=\hsize}X 
         >{\hsize=0.7\hsize\linewidth=\hsize}X}
        \toprule
        \textbf{Key} & \textbf{Quality} & \textbf{Description} \\
        \midrule
        NFR1  & Accessibility    & User must be able to access and retrieve data \\
        NFR2  & Authenticity     & Source of identity data must be trustworthy and provable \\
        NFR3  & Autonomy         & User must be able to manage their identity independently \\
        NFR4  & Availability     & Identity data must be available at any time \\
        NFR5  & Compatibility    & Identity data must be compatible with legacy systems \\
        NFR6  & Consent          & User must explicitly consent to the use of their data \\
        NFR7  & Control          & User must be able to control access to their identity data \\
        NFR8  & Cost             & All components must have minimal costs \\
        NFR9  & Decentralization & All components should not rely on centralized elements \\
        NFR10 & Existence        & User identity must have an independent existence without relying on other services \\
        NFR11 & Interoperability & Identity data must be usable across different platforms and services \\
        NFR12 & Persistence      & Identity data must remain valid and accessible for as long as necessary \\
        NFR13 & Portability      & User must be able to move their identity data \\
        NFR14 & Privacy          & User must be able to minimize information required to share \\
        NFR15 & Protection       & Identity data must be protected against misuse \\
        NFR16 & Recoverability   & User must be able to recover identity data in case of loss and compromise \\
        NFR17 & Representation   & Users must be able to create multiple identities \\
        NFR18 & Security         & All components must ensure the data is secure \\
        NFR19 & Single Source    & User must be the single authoritative source of their identity \\
        NFR20 & Standard         & Credentials must adhere to open standards \\
        NFR21 & Transparency     & Information about data use must be readily available \\
        NFR22 & Usability        & User must be able to use their data efficiently and intuitively \\
        NFR23 & User Experience  & Identity management process must be simple, consistent, and user-friendly \\
        NFR24 & Verifiability    & Identity data must be verifiable \\
        \bottomrule
    \end{tabularx}
\end{table}

\section{Operationalization Methodology}\label{sec:methodology}

    The operationalization of requirements presented in this work builds on the requirements categorization introduced in \cite{paper_categorization} and draws inspiration from the Tropos methodology \cite{318,319,320}.  
    The applicability of the Tropos methodology is motivated by the need to model the relationships among all involved stakeholders (or actors) within the DI/SSI systems. 
    Meanwhile, previously identified responsibilities and ownerships in \cite{paper_categorization} enable differentiation between functional and non-functional dependencies within a system. 

    Operationalization of NFR followed a four-step methodology, building on the generalized use case describing DI/SSI system functionalities, as presented in \cite{paper_categorization}. 
    First, the capabilities of each actor (\textit{i.e.}, data owner, issuer, verifier) are outlined and mapped to design patterns from \cite{292,294}.
    Second, a functional model was developed for each actor based on identified capabilities. 
    Third, predicates and axioms were formulated to formalize the functional model and support reasoning about the capabilities of actors and interactions within the system. 
    Lastly, a set of FR was formulated for each NFR. 

    \paragraph{\textbf{Capabilities and Constraints.}}
        As the first step, building on design patterns \cite{292,294}, relevant literature (\textit{e.g.}, \cite{17}), and NFR categorization \cite{paper_categorization}, capabilities and constraints were identified. 
        Each design pattern, as presented in \cite{292,294}, was closely considered, focusing on benefits and functionality (\textit{i.e.}, which actor of a system benefits from the pattern and implied functionality). 
        Based on the initial analysis of the design patterns, NFR categorization \cite{paper_categorization} was extended by identifying actors' capabilities. 
        Each capability was formulated into a capability statement which ``expresses a capability required by one or more identified” actors \cite[p.84]{334}.
        Statements follow the format specified in \cite{334}: the <actor> shall be able to <capability>.

        \cite{paper_categorization} identified several requirements as constraints, namely cost (NFR8), decentralization (NFR9), portability (NFR13), security (NFR18), usability (NFR22), and user experience (NFR23). 
        Considering identified capabilities and the constraint definition (it does not add any capability to a system), compatibility (NFR5) and standard (NFR20) were added as constraints, while portability (NFR13) was removed. 

    \paragraph{\textbf{Functional Model.}}
        As the second step, functional dependencies between actors were formulated and a functional model developed.
        Functional model illustrates ``how the system functions are performed” \cite[p.69]{333}.
        The main difference between the functional model and the dependency model proposed in \cite{paper_categorization} is that the former captures logical relationships among actors, whereas the latter describes how these relationships can be realized. 
        To verify the functional model, identified capabilities were mapped to the design patterns presented in \cite{292,294}.
        The mapping was done in two steps. 
        First, all design patterns were assigned to a specific NFR, regardless of the actor's viewpoint. 
        Second, irrelevant patterns for each actor were removed, leaving only relevant patterns from the actor's viewpoint. 
        

    \paragraph{\textbf{Predicates and Axioms.}}\label{sec:methodology_predicates}

        To enable formal representation, the next step formalized the functional model with logic-based predicates and axioms. 
        In software engineering, a predicate is a logical expression that represents an observable behavior and helps to decide whether it satisfies a specification with a definite ``yes” or ``no” \cite{324}.
        In other words, a predicate represents a condition that takes an input and determines whether this condition is satisfied. 
        Predicate expression is a logical expression built using primitive predicates (basic, representing a set of simple tuples), logical operations (AND, OR, NOT, IMPLIES), quantifiers ($\forall$ and $\exists$), and parentheses \cite{324}.
        Predicates show possible interactions between actors ($a$) and services ($s$). 



        Axioms were derived from the outlined predicates, which provide a systematic method for modeling and enable the drawing of correct conclusions from an intuitive approach \cite{319}.
        In this work, service ($s$) refers to functionality, such as credential generation, verification, revocation, storage, and retrieval, while data ($d$) refers to credential or personal data that the data owner stores, retrieves, and shares for verification with verifiers. 
        In some cases, it is necessary to indicate the relationships among various actors. 
        A generic actor ($a$) used in predicates is replaced with a specific actor, such as a data owner ($o$), verifier ($v$), or issuer ($i$). 
        Services are fulfilled by different actors who own them. 
        For instance, a verifier has a verification service (for a credential and the issuer's signature), and an issuer has credential issuance and revocation services. 


    \paragraph{\textbf{Formulation of Functional Requirements.}}

        As the final step, the final set of FR is derived and formulated. 
        While previous steps provide an overview of FR, it is important to summarize FR for each NFR in a common, unambiguous format to facilitate communication between actors and enhance the quality of requirements.
        Methodologically developed boilerplates for formulating requirements help minimize semantic errors and structure requirements in a common pattern. 
        A methodology for requirements formulation outlined in \cite{298} was used in this step because it is the only approach that explicitly considers legal requirements. 
        Legal requirements are necessary in the DI/SSI domain, as an increasing number of regulations are becoming mandatory across countries and are essential for real-world adoption. 
        Therefore, a system must embed regulations, and FR should reflect the legal implications.

        The formulation of FR was achieved in five steps. 
        The first step identified a legal obligation for each NFR drawn from Electronic Identification, Authentication and Trust Services (eIDAS) and the General Data Protection Regulation (GDPR).  
        The second step identified a process or method for achieving each NFR, based on an evaluation of the system's goals or capabilities \cite{298,325}. 
        In this work, a process or method corresponds to the capabilities of actors. 
        The third step classified the system activity as (i) autonomous system activity, (ii) user interaction, or (iii) interface requirement (see Table \ref{tab:system_activity}) \cite{298}.
        As the fourth step, functional and constraint statements were formulated. 
        Lastly, missing details, such as objects, conditions, or actions, and logical and temporal conditions under which a requirement was considered fulfilled, were added. 

        \begin{table}[h]
    \centering
    \caption{System Activities}
    \label{tab:system_activity}
    \def\arraystretch{1.5}%
    \footnotesize
    \rowcolors{2}{white}{gray!10}
    \begin{tabularx}{\textwidth}
        {>{\hsize=0.1\hsize\linewidth=\hsize}X 
         >{\hsize=0.2\hsize\linewidth=\hsize}X 
         >{\hsize=0.35\hsize\linewidth=\hsize}X
         >{\hsize=0.35\hsize\linewidth=\hsize}X
         }
        \toprule
        \textbf{Type} & \textbf{Name} & \textbf{Description} & \textbf{Template} \\
        \midrule
        
        T1 & Autonomous System Activity & Describes autonomous system activities. The user is not part of the activity. & The system [shall, should, will, may] $<$process verb$>$. \\

        T2 & User Interaction & The system provides functionality to a user or interacts directly with a user. & The system [shall, should, will, may] provide $<$whom?$>$ with the ability to $<$process verb$>$. \\

        T3 & Interface Requirement & The system activity depends on other systems. & The system [shall, should, will, may] be able to $<$process verb$>$. \\

        \bottomrule
    \end{tabularx}
\end{table}
\section{Operationalization Results}\label{sec:results}

    Following the operationalization methodology outlined in the previous section, a set of FR was derived.
    Each methodology step produced a separate output, detailed in this section, which supports the derivation and formulation of FR. 
    
    \subsection{Capabilities and Constraints}\label{sec:results_capabilities}

        As a result of the first step in methodology, a list of capabilities was formulated.
        Table \ref{tab:capabilities} outlines the identified capabilities for each actor.
        The following requirements were identified as constraints: compatibility (NFR5), cost (NFR8), decentralization (NFR9), security (NFR18), standard (NFR20), usability (NFR22), and user experience (NFR23). 
        
        \begin{table}[]
    \centering
    \caption{Capabilities of Actors}
    \label{tab:capabilities}
    \def\arraystretch{1.5}%
    \footnotesize
    \rowcolors{2}{white}{gray!10}
    \begin{tabularx}{\textwidth}
        {>{\hsize=0.2\hsize\linewidth=\hsize}X 
         >{\hsize=0.05\hsize\linewidth=\hsize}X 
         >{\hsize=0.75\hsize\linewidth=\hsize}X}
        \toprule
        \textbf{Actor} & \textbf{\#} & \textbf{Capability} \\
        \midrule
        Data Owner & 1  & Shall be able to hold personal data \\
                   & 2  & Shall be able to present personal data to the issuer \\
                   & 3  & Shall be able to request a credential from the issuer \\
                   & 4  & Shall be able to generate a new DID \\
                   & 5  & Shall be able to generate keys \\
                   & 6  & Shall be able to store a credential pr personal data \\
                   & 7  & Shall be able to export a credential or personal data \\
                   & 8  & Shall be able to import a credential or personal data \\
                   & 9  & Shall be able to retrieve a credential or personal data \\
                   & 10 & Shall be able to present a credential for verification \\
                   & 11 & Shall be able to request a service from the verifier \\
        Verifier   & 12 & Shall be able to offer a service \\
                   & 13 & Shall be able to verify a credential of the data owner \\
                   & 14 & Shall be able to verify a signature of the issuer \\
        Issuer     & 15 & Shall be able to issue a credential \\
                   & 16 & Shall be able to revoke a credential \\
                   & 17 & Shall be able to register with the registry \\
        System     & 18 & Shall be able to keep the issuer registry \\
                   & 19 & Shall be able to keep the revocation registry \\
                   & 20 & Shall be able to keep the schema registry \\
        Wallet     & 21 & Shall be able to store data \\
                   & 22 & Shall be able to support data export and import \\
                   & 23 & Shall be able to recover access to data \\
                   & 24 & Shall be able to provide an interface for interactions \\
        \bottomrule
    \end{tabularx}
    \vspace{-0.3cm}
\end{table}

    \subsection{Functional Model}\label{sec:results_functional_model}

\begin{figure}[]
            \centering
            \includegraphics[keepaspectratio=true, width=0.8\linewidth]{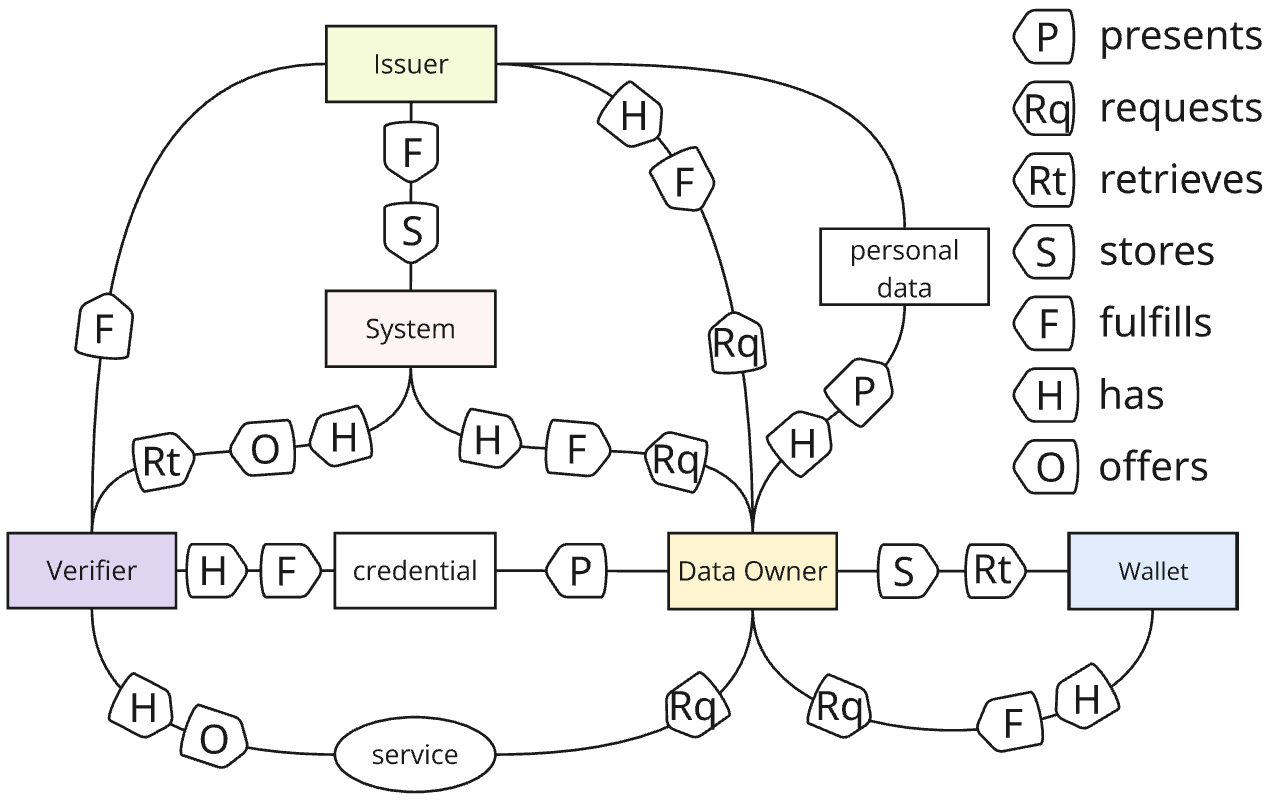}
            \caption{Functional Model}
            \label{fig:functional_model}
        \end{figure}

\begin{figure}[]
            \centering
            \includegraphics[keepaspectratio=true, width=\linewidth]{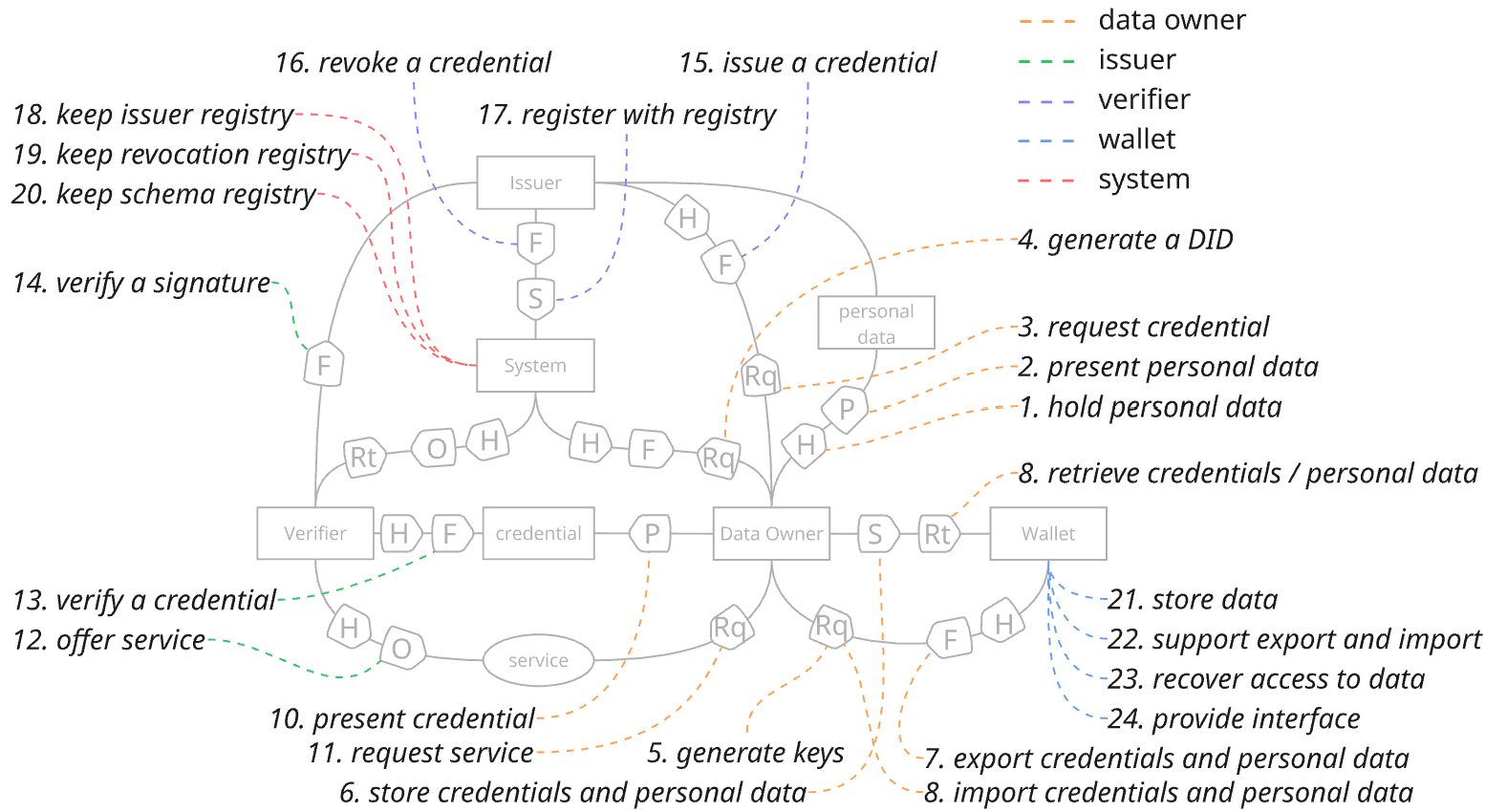}
            \caption{Actor Capabilities}
            \label{fig:capabilities}
        \end{figure}

        Figure \ref{fig:functional_model} illustrates the functional model of DI/SSI systems based on identified capabilities (see Table \ref{tab:capabilities}). 
        In Figure \ref{fig:functional_model}, presentation of a credential labeled as $P$, request for service or data as $Rq$, retrieval of credential or data as $Rt$, storage of credential or data as $S$, fulfillment of requested service (\textit{e.g.}, credential issuance and verification, decentralized identifier (DID) creation) as $F$, having a service as $H$, and offering a service as $O$. 
        Based on the functional model and capabilities summarized in the previous step, Figure \ref{fig:capabilities} summarizes the capabilities of each actor. 
        An extensive mapping between capabilities and NFR (outlined in Table \ref{tab:nfr}) was performed. 
        Appendix A\footnote{https://github.com/schummd/operationalization} provides the complete tables and figures for each actor's capabilities and constraints. 
        For simplicity, part of the capabilities boilerplate (``shall be able to'') was omitted.
        
        The detailed results of functional model verification and mapping between capabilities and design patterns are presented in Appendix B\footnote{https://github.com/schummd/operationalization}. 
        Verification identified a limitation of the existing design patterns (\textit{i.e.}, \cite{292,294}). 
        That is, existing design patterns do not distinguish between actors and their interactions, leading some patterns not to fit a particular capability, even when the pattern exists for that NFR.
        For example, the capability of the data owner to present personal data to the issuer (\#2 in Table \ref{tab:capabilities}) maps to authenticity NFR, but there are no relevant patterns, because they refer strictly to the issuance of a verifiable credential (VC) within DI/SSI interactions.
        However, the presentation of personal data to request credential issuance predates the use of DID or VC and is not considered in the existing patterns. 
        Additionally, several capabilities and NFR are not represented in any design patterns, such as consent, existence, and autonomy. 
                
        \vspace{-10px}

    \subsection{Predicates and Axioms}

        \begin{table}[h]
    \centering
    \caption{Predicates for Functional Model}
    \label{tab:predicates}
    \def\arraystretch{1.5}%
    \footnotesize
    \begin{tabular}{ l l }
        \toprule
        \textbf{Functional Model} & \textbf{Pattern} \\
        \midrule
        
        \rowcolor{gray!10}
        Ownership & owns(Actor: $a$, Service: $s$) \\
        \rowcolor{gray!10}
         & has(Actor: $a$, Service: $s$) \\

        Responsibility & fulfills(Actor: $a$, Service: $s$) \\
         & offers(Actor: $a$, Service: $s$) \\

        \rowcolor{gray!10}
        Execution & requests(Actor: $a$, Service: $s$) \\
        \rowcolor{gray!10}
         & presents(Actor: $a$, Service: $s$) \\
         \rowcolor{gray!10}
         & retrieves(Actor: $a$, Service: $s$) \\
         \rowcolor{gray!10}
         & stores(Actor: $a$, Service: $s$) \\
        
        \bottomrule
    \end{tabular}
\end{table}
        \begin{table}[!htbp]
    \centering
    \caption{Axioms for Functional Model}
    \label{tab:axioms}
    \def\arraystretch{1.5}%
    \footnotesize
    \rowcolors{2}{white}{gray!10}
    \begin{tabular}{ l p{0.45\linewidth} p{0.42\linewidth} }
        \toprule
        \textbf{\#} & \textbf{Axiom} & \textbf{Description} \\
        \midrule
        
        Ax1 & $has(a,[s,d]) \leftarrow owns(a,[s,d])$ 
        & If an actor owns a service or data, they have it. \\

        Ax2 & $presents(a,d) \leftarrow has(a,d)$ 
        & If an actor has data, they present it. \\

        Ax3 & $fulfills(a,s) \leftarrow has(a,s) \land offers(a,s)$ 
        & If an actor has and offers a service, they fulfill it. \\

        Ax4 & $stores(a,d) \leftarrow has(a,d)$ 
        & If an actor has data, they store it. \\

        Ax5 & $retrieves(a,d) \leftarrow stores(a,d)$ 
        & If an actor stores data, they retrieve it. \\

        Ax6 & $retrieves(a,d) \leftarrow stores(b,d) \land (has(c,d) \land offers(c,d))$ 
        & If actor $b$ stores data and actor $c$ (system) has and offers the data, actor $a$ retrieves it. \\

        Ax7 & $fulfills(a,s) \leftarrow (presents(b,d) \lor retrieves(a,d)) \land requests(b,s)$ 
        & If actor $b$ presents or retrieves data and requests a service, actor $a$ fulfills the service. \\

        \bottomrule
    \end{tabular}
\end{table}

        The set of predicates outlined in Table \ref{tab:predicates} describes the functional model and comprises three sub-models. 
        The ownership model ($o$) includes the $own(a, s)$ predicate, which represents the actor that owns a service and is fulfilled if the actor has a capability to perform the service that he offers. 
        Another predicate has$(a, s)$ shows that an actor who has a service has an authority over the use of that service.
        The responsibility model ($r$) includes two predicates. 
        The fulfills$(a, s)$ predicate is true when a service is provided by an actor, and the offers$(a, s)$ predicate indicates that an actor offers a service when he has the capabilities to fulfill the service. 
        The execution model ($e$) contains four predicates.
        The requests$(a, s)$ predicate holds when the actor wants a service to be fulfilled, and the provides$(a, s)$ predicate is satisfied when the actor has a capability to fulfill the service. 
        The presents$(a, b)$ predicate holds when actor $a$ presents a credential or necessary data to actor $b$. 
        The retrieves$(a, s)$ predicate is fulfilled when actor $a$ receives the data from the service. 
        The stores$(a, s)$ hold when an actor successfully stores the data at the service. 
        Table \ref{tab:axioms} outlines axioms for the DI/SSI domain.

    \subsection{Formulating Functional Requirements}\label{sec:res_formulating_fr}
        
        Building on the previous steps, a set of FR was developed. 
        For each NFR, the description (\textit{e.g.}, ownership, responsibility, functional process, legal requirements, system activity, and constraints), functional statements, and constraint statements are presented in three tables. 
        Section \ref{sec:results_template} outlines the table templates, and Section \ref{sec:results_example} shows an example of consent (NFR6). 
        It is assumed that processes are performed through a compatible device available to each actor, and that processes involving credential or signature verification, as well as credential revocation, rely on cryptographic validity. 
        The remaining NFR and their corresponding FR are detailed in Appendix C\footnote{https://github.com/schummd/operationalization}. 
        The following actor definitions are used in this work: 
        \begin{itemize}
            \item The \textit{data owner} ($o$; or identity owner, DID subject) is the entity that receives, stores, and presents a credential (or VC). 
            \item The \textit{issuer} ($i$) creates and signs credentials, asserting claims about the data owner.
            \item The \textit{verifier} ($v$) requests and verifies the validity of credentials presented by the data owner. 
            \item A \textit{global system} ($s$) represents a blockchain component that enables decentralization of the identity management system.
            \item A \textit{local system}, referred to as a wallet ($w$), is software that the data owner interacts with. 
        \end{itemize}
        
        \subsubsection{Functional Requirements Template.}\label{sec:results_template}
            Functional statements are formulated using an established boilerplate from \cite{298}, which employs a five-step process, and use the template presented in Table \ref{tab:statements_template}. 
            The process includes outlining (i) legal obligations, (ii) functionality (notated as <process>), (iii) system activity (autonomous, user interaction, interface requirement), (iv) suitable axioms, (v) objects, and (vi) constraints \cite{298}.
            The results of this process are summarized in the NFR description table, as shown in Table \ref{tab:nfr_template}.
            Conditions, which follow the template in Table \ref{tab:conditions_template}, are necessary for the complete satisfaction of a specified FR. 
    
            \begin{table}[h]
    \centering
    \caption{ Non-Functional Requirement Description Template }
    \label{tab:nfr_template}
    \def\arraystretch{1.5}%
    \footnotesize
    \rowcolors{2}{white}{gray!10}
    \begin{tabularx}{\textwidth}
        {>{\hsize=0.25\hsize\linewidth=\hsize}X 
         >{\hsize=0.75\hsize\linewidth=\hsize}X}
        \toprule
        Key               & NFR X \\
        NFR               & Name \\
        Owner             & [ $o$, $v$, $i$, $s$, $w$ ] \\
        Responsibility    & [ $o$, $v$, $i$, $s$, $w$ ] \\
        Legal Requirement & [GDPR Article X, eIDAS] \\
        Process or Method & [ $o$, $v$, $i$, $s$, $w$ ]: Capability \# \\
        Axioms            & [ Ax1, Ax2, Ax3, Ax4, Ax5, Ax6, Ax7 ] \\
        System Activity   & [ T1, T2, T3 ] \\
        Constraints       & [ NFR5, NFR8, NFR9, NFR13, NFR18, NFR20, NFR22, NFR23 ] \\
        \bottomrule
    \end{tabularx}
\end{table}
\vspace{-25px}

\begin{table}[!htbp]
    \centering
    \caption{ Functional Statements Template }
    \label{tab:statements_template}
    \def\arraystretch{1.5}%
    \footnotesize
    \rowcolors{2}{white}{gray!10}
    \begin{threeparttable}
        \begin{tabularx}{\textwidth}
        {>{\hsize=0.12\hsize\linewidth=\hsize}X 
         >{\hsize=0.88\hsize\linewidth=\hsize}X}
            \toprule
            \textbf{Key} & \textbf{Statements} \\
            \midrule
            FR X.X\tnote{a} & THE SYSTEM shall PROVIDE [ $o$, $v$, $i$, $s$, $w$ ] WITH THE ABILITY TO ... \\
                            & THE SYSTEM shall ... \\
            \bottomrule
        \end{tabularx}
        \begin{tablenotes}
            \item[a] Follows the naming of NFR. For example, a FR for NFR1 is FR1.1.  
        \end{tablenotes}
    \end{threeparttable}
\end{table}
\vspace{-25px}

\begin{table}[!htbp]
    \centering
    \caption{ Functional Conditions Template }
    \label{tab:conditions_template}
    \def\arraystretch{1.5}%
    \footnotesize
    \rowcolors{2}{white}{gray!10}
    \begin{threeparttable}
        \begin{tabularx}{\textwidth}
        {>{\hsize=0.12\hsize\linewidth=\hsize}X 
         >{\hsize=0.88\hsize\linewidth=\hsize}X}
            \toprule
            \textbf{Key} & \textbf{Statements} \\
            \midrule
            C X.X.X\tnote{a} & \textit{No specific template} \\
            \bottomrule
        \end{tabularx}
        \begin{tablenotes}
            \item[a] Follows the naming of FR. For example, a condition for FR1.1 is C1.1.1.  
        \end{tablenotes}
    \end{threeparttable}
\end{table}
        
        \subsubsection{Functional Requirements Example.}\label{sec:results_example}
            An example of consent (NFR6) is shown in this section to illustrate how previously outlined templates are used.
            Consent is a suitable example because it encompasses processes involving multiple actors, uses multiple functional statement templates, and references several legal regulations.     
            For consent (NFR6), the data owner is both the owner and key actor in fulfilling this requirement (see Table \ref{tab:nfr6_details}). 
            The issuer and verifier also share responsibility, as they must obtain and maintain the data owner's consent before processing personal data or credentials \cite{paper_categorization}. 
            
            Several GDPR provisions specify the legal requirements for consent, such as Article 6(1)(a), Article 7, and Recital 23. 
            According to Article 6(1)(a) \cite{337}, the lawful processing of personal data requires that the data owner provide an explicit consent for a specific purpose. 
            Article 7 \cite{338} further establishes the conditions under which consent must be obtained and documented, including the ability to withdraw consent easily. 
            Recital 23 \cite{339} emphasizes that data owners must be fully informed about the intended use of their data, and that consent must be an unambiguous indication of agreement, typically through a clear affirmative action.
    
            The processes linked to consent include presenting personal data to the issuer (capability \#2), requesting a credential from the issuer (capability \#3), and presenting credentials for verification (capability \#10).
            However, these actions should not occur in isolation, but must be preceded by a clear, user-centric mechanism that explains the purpose of the data use and obtains the user’s agreement. 
            Thus, the consent requirement focuses not only on the act of data presentation but also on ensuring transparent communication and the collection of explicit consent before data sharing.

            \begin{table}[]
    \centering
    \caption{ Consent NFR Description }
    \label{tab:nfr6_details}
    \def\arraystretch{1.5}%
    \footnotesize
    \rowcolors{2}{white}{gray!10}
    \begin{tabularx}{\textwidth}
        {>{\hsize=0.25\hsize\linewidth=\hsize}X 
         >{\hsize=0.75\hsize\linewidth=\hsize}X}
        \toprule
        Key               & NFR6 \\
        NFR               & Consent \\
        Owner             & $o$ \\
        Responsibility    & $o$, $v$, $i$ \\
        Legal Requirement & GDPR Article 6(1)(a): Lawfulness of Processing \cite{337} \\ 
                          & GDPR Article 7: Conditions for Consent \cite{338} \\ 
                          & GDPR Recital 23 \cite{339} \\
        Process or Method & $o$: Present personal data to the issuer (\#2) \\
                          & $o$: Request credential from the issuer (\#3) \\
                          & $o$: Present credential or proof for verification (\#10) \\
        Axioms            & Ax2, Ax3 \\
        System Activity   & T1, T2, T3 \\
        Constraints       & NFR8 \\
        \bottomrule
    \end{tabularx}
\end{table}

            \begin{table}[]
    \centering
    \caption{ Consent Functional Statements }
    \label{tab:nfr6_fr}
    \def\arraystretch{1.5}%
    \footnotesize
    \rowcolors{2}{white}{gray!10}
    \begin{tabularx}{\textwidth}
        {>{\hsize=0.1\hsize\linewidth=\hsize}X 
         >{\hsize=0.9\hsize\linewidth=\hsize}X}
        \toprule
        \textbf{Key} & \textbf{Statements} \\
        \midrule
        FR6.1 & THE SYSTEM shall inform the $o$ about the intended use of the credential. \\
        FR6.2 & THE SYSTEM shall PROVIDE the $o$ WITH THE ABILITY TO consent to credential or personal data processing. \\
        FR6.3 & THE SYSTEM shall PROVIDE the $o$ WITH THE ABILITY TO withdraw consent to credential or personal data processing. \\
        FR6.4 & THE SYSTEM shall collect the $o$ agreement to credential or personal data processing. \\
        FR6.5 & THE SYSTEM shall use unambiguous language to obtain consent. \\
        FR6.6 & THE SYSTEM shall PROVIDE the $o$ WITH THE ABILITY TO present personal data to the issuer. \\
        FR6.7 & THE SYSTEM shall PROVIDE the $o$ WITH THE ABILITY TO request a credential from the issuer. \\
        FR6.8 & THE SYSTEM shall PROVIDE the $o$ WITH THE ABILITY TO present a credential for verification. \\
        FR6.9 & THE SYSTEM shall PROVIDE the $o$ WITH THE ABILITY TO share only a proof of a credential. \\
        \bottomrule
    \end{tabularx}
\end{table}

            Functional statements, outlined in Table \ref{tab:nfr6_fr}, comprehensively capture the operationalization of this requirement. 
            The predicates and axioms, outlined in Section \ref{sec:methodology_predicates}, serve as the fundamental relationships and inference rules on which FR are based.
            Functional statements are derived by integrating relevant processes (\textit{i.e.}, capabilities) and supplementing them with axioms to highlight the implications (\textit{e.g.}, $o$ owning a credential implies the ability to present).
            As a result of integration, functional statements (\textit{e.g.}, FR6.1 and FR6.4) ensure that the consent process is not merely procedural, but reflects the decision-making process of the data owner, as well as supports personal data or credentials presentation, and credentials requests, which depend on prior user consent (\textit{e.g.}, FR6.7 and FRF6.9). 
            These steps reflect user interaction, autonomous system activity, and interface requirement types of system activity, illustrating how consent integrates across the system and user layers.
            
            The conditions for fulfilling consent requirement (see Table \ref{tab:nfr6_conditions}) include mechanisms to inform the data owner and to capture, store, and manage their consent and its withdrawal.
            A key constraint is cost (NFR8), as implementing robust, user-friendly consent workflows and ensuring compliance with legal standards introduces technical and operational overhead.
            As a result, these FR ensure that all processing of personal data and credentials within the system is lawful, transparent, and well-informed.
            
            \begin{table}[h]
    \centering
    \caption{ Consent Functional Conditions }
    \label{tab:nfr6_conditions}
    \def\arraystretch{1.5}%
    \footnotesize
    \rowcolors{2}{white}{gray!10}
    \begin{tabularx}{\textwidth}
        {>{\hsize=0.07\hsize\linewidth=\hsize}X 
         >{\hsize=0.93\hsize\linewidth=\hsize}X}
        \toprule
        \textbf{Key} & \textbf{Statements} \\
        \midrule
        C6.1.1 & The $s$ collected information from the $v$ on how the credential or personal data will be used. \\
        C6.2.1 & The credential has not yet been shared. \\
        C6.3.1 & The consent has been previously given. \\
        C6.4.1 & The $o$ agrees to data processing before sharing credentials. \\
        C6.5.1 & The $o$ can be in a different age category and have different technical knowledge. \\
        C6.6.1 & The $o$ is aware of how credentials or personal data are used by the $v$. \\
        C6.6.2 & The $i$ informed the $o$ what personal data is required. \\
        C6.7.1 & The $o$ is aware of what personal data will be used for credential issuance. \\
        C6.8.1 & The $o$ provided consent to credential verification. \\
        C6.8.2 & Each presented credential is treated separately and requires its own consent. \\
        C6.8.3 & Verification requests are triggered by the $o$. \\
        C6.8.4 & The $o$ selects which attributes to share before presentation. \\
        C6.8.5 & The presented credential includes necessary metadata. \\
        C6.9.1 & The $o$ can generate a proof. \\
        \bottomrule
    \end{tabularx}
    \vspace{-15px}
\end{table}

        \subsubsection{Functional Requirements Validation.}\label{sec:results_validation}
            As a result of the previous steps, a set of 39 FR was formulated.
            To validate the FR, duplicates were first removed. 
            Then, based on validation criteria for the set of requirements and for each requirement outlined in \cite{334}, FR were validated and adjusted accordingly. 
            Table \ref{tab:set_criteria} outlines validation criteria for the set of requirements and Table \ref{tab:each_criteria} outlines validation criteria for each requirement.
            \vspace{-4px}
            \paragraph{\textbf{Validating Set of Requirements.}}
                The set of requirements is \textit{complete} since it covers the main actors of the DI/SSI system, their capabilities, and responsibilities, supported by existing literature (\textit{e.g.}, \cite{17}) and prior work \cite{paper_categorization}.
                Legal regulations were considered, extending FR to reflect these legal requirements and ensure comprehensive coverage. 
                The set of requirements is \textit{non-redundant} and \textit{modular}, as all duplicates were removed and the remaining unique functional statements were arranged by the topics they address. 
                Each functional statement has a unique identifier. 
                The set is \textit{consistent} because every requirement addresses a unique functionality of the system.
                The set is \textit{structured} because every functional statement closely follows an established boilerplate.
                Lastly, the set of requirements is \textit{traceable} since every statement links to previously identified actors, their capabilities, and corresponding NFR.
                Therefore, the set of requirements satisfies all of the criteria for a set of statements. 
    
                \begin{table}[h]
    \centering
    \caption{ Criteria for a Set of Requirement \cite{334} }
    \label{tab:set_criteria}
    \def\arraystretch{1.5}%
    \footnotesize
    \rowcolors{2}{white}{gray!10}
    \begin{tabular}{ l p{0.76\linewidth} }
        \toprule
        \textbf{Criteria} & \textbf{Description} \\
        \midrule
        Complete      & All requirements are present. \\
        Consistent    & No two requirements are in conflict. \\
        Non-redundant & Each requirement is expressed once. \\
        Modular       & Requirements that belong together are close to one another. \\
        Structured    & The requirements document has a clear structure. \\
        Satisfied     & The appropriate degree of traceability coverage has been achieved. \\
        Qualified     & The appropriate degree of traceability coverage has been achieved. \\
        \bottomrule
    \end{tabular}
\end{table}
\vspace{-20px}

\begin{table*}[h]
    \centering
    \caption{ Criteria for Each Requirement Statement \cite{334} }
    \label{tab:each_criteria}
    \def\arraystretch{1.5}%
    \footnotesize
    \rowcolors{2}{white}{gray!10}
    \begin{tabular}{ p{0.18\linewidth} p{0.76\linewidth} }
        \toprule
        \textbf{Criteria} & \textbf{Description} \\
        \midrule
        Atomic     & Each statement carries a single traceable element. \\
        Unique     & Each statement can be uniquely identified. \\
        Feasible   & Technically possible within cost and schedule. \\
        Legal      & Legally possible. \\
        Clear      & Each statement is clearly understandable. \\
        Precise    & Each statement is precise and concise. \\
        Verifiable & Each statement is verifiable and it is known how. \\
        Abstract   & Does not impose a solution or design specific to the layer below. \\
        \bottomrule
    \end{tabular}
\end{table*}
    
            \paragraph{\textbf{Validating Each Requirement.}}
                Each of the 39 FR was evaluated against criteria for the individual requirements, outlined in Table \ref{tab:each_criteria}.
                Some of the FR did not meet or only partially addressed the criteria.
                Particularly, criteria such as \textit{atomic}, \textit{clear}, and \textit{precise} were frequently unaddressed.  
                To ensure all criteria are satisfied, functional statements were modified accordingly. 
                Table \ref{tab:nfr6_updated_fr} shows an output of the updated functional statement for consent (previously outlined in Table \ref{tab:nfr6_fr}). 
                The statements were modified according to the validation criteria, specifying which criterion was addressed by the modification. 
                Multiple FR were split into two separate FR to achieve the atomic requirement. 
                To indicate the split, the key for each remains as the original, but was extended with a letter (\textit{e.g.}, FR8 became FR8a and FR8b). 
                Additionally, the unique identifier was updated for each FR. 
                The entire updated set of FR is provided in Appendix D\footnote{https://github.com/schummd/operationalization}. 
                
                \begin{table}[!htbp]
    \centering
    \caption{ Updated Functional Statements for Consent }
    \label{tab:nfr6_updated_fr}
    \def\arraystretch{1.5}%
    \footnotesize
    \rowcolors{2}{white}{gray!10}
    \begin{tabular}{ l p{0.6\linewidth} p{0.12\linewidth} l }
        \toprule
        \textbf{Key} & \textbf{Updated Functional Statements} & \textbf{Criteria} & \textbf{Trace}\\
        \midrule
        FR18 & THE SYSTEM shall obtain consent of $o$ using language that meets a predefined readability standard (\textit{e.g.}, Flesch-Kincaid Grade 8 or lower). & Verifiable & FR6.5 \\
        
        FR32 & THE SYSTEM shall inform the $o$ about the purpose, scope, and recipients of the data usage through an accessible notification or consent screen. & Clear & FR6.1 \\
        
        FR33a & THE SYSTEM shall obtain explicit consent from the $o$ for credential processing via a user interface prompt before any processing occurs. & Clear \newline Atomic & FR6.4 \\
        FR33b & THE SYSTEM shall obtain explicit consent from the $o$ for personal data processing via a user interface prompt before any processing occurs. & Clear \newline Atomic & FR6.4 \\

        FR44 & THE SYSTEM shall PROVIDE the $o$ WITH THE ABILITY TO request a credential from a selected $i$. & Clear & FR6.7 \\

        FR46 & THE SYSTEM shall PROVIDE the $o$ WITH THE ABILITY TO give informed consent before any processing of credentials or personal data, including issuance, sharing, or verification. & Clear & FR6.2 \\

        FR47 & THE SYSTEM shall PROVIDE the $o$ WITH THE ABILITY TO withdraw previously given consent for the processing of credential or personal data. & Clear & FR6.3 \\

        FR52 & THE SYSTEM shall PROVIDE the $o$ WITH THE ABILITY TO present selected personal data to an $i$ upon request. & Clear & FR6.6 \\

        FR53 & THE SYSTEM shall PROVIDE the $o$ WITH THE ABILITY TO present a cryptographically verifiable credential to a $v$ upon request. & Clear & FR6.8 \\

        FR54 & THE SYSTEM shall PROVIDE the $o$ WITH THE ABILITY TO present a cryptographically verifiable proof of a credential to a $v$ upon request. & Clear & FR6.9 \\
        
        \bottomrule
    \end{tabular}
\end{table}

            
\section{Conclusions and Future Work}\label{sec:conclusions}

    This work presented a systematic approach for operationalizing NFR into FR within DI/SSI systems. 
    The operationalization of NFR develops a comprehensive, but not exhaustive, set of FR for a generalized DI/SSI use case, which covers the core operations of the DI/SSI system. 
    The methodology consists of four steps, namely (i) deriving capabilities and constraints of each actor, (ii) describing and (iii) formalizing the functional model, and (iv) formulating FR. 
    Together, these steps provide a traceable transformation from high-level quality requirements (NFR) to system operations (FR). 
    The result ensures that the resulting FR are traceable and consistent within the underlying system architecture and actor responsibilities. 
    The formal notations of the functional model establish a formal foundation that supports reasoning about actor capabilities and interactions within the system. 
    Overall, this work presented the first step toward building a transparent and reproducible evaluation approach for DI/SSI systems.

    It is important to note the role of decentralization in DI/SSI systems. 
    Considering the definitions of DI and SSI, as presented in \cite{survey, chapter}, the underlying difference between the two concepts is the decentralization of governance. 
    That is, no single element of a system should be controlled by a single organization. 
    There are no direct FR that satisfy these requirements, because the difference is not a system function but constraints on relevant functionalities. 
    Based on the functional model presented in this paper (Section \ref{sec:results_functional_model}), both DI and SSI have the same underlying functionality. 
    However, for a DI to be an SSI, decentralization constraints on every applicable FR must be met. 
    The complete table summarizing the decentralization constraints is provided in Appendix E\footnote{}.

    The future work will formalize the FR introduced in this work and design an evaluation framework to verify requirement satisfaction, detect violations, and assess requirement coverage across use cases.
    Particularly, the future work can focus on the verification of decentralization requirements, as an underlying difference between DI and SSI systems. 
    Additionally, Section \ref{sec:results_functional_model} identified several gaps in the existing design patterns of DI/SSI systems. 
    Future work on design patterns can address capabilities and NFR that did not map to the existing design patterns, such as consent, autonomy, and existence.







\bibliographystyle{splncs04}
\bibliography{ref}

@unpublished{paper_categorization,
    author = {Daria Schumm and Burkhard Stiller},
    title = {{Rethinking Self-Sovereign Identity Principles: An Actor-Oriented Categorization of Non-Functional Requirements}},
    year = {2026},
    note = {{Under Review. Preprint is available at: \url{https://arxiv.org/pdf/2603.23177}}}
}

@incollection{chapter,
    author = {Daria Schumm and Bruno Rodrigues and Burkhard Stiller},
    editor = {Lachlan Robb and John Flood}, 
    title = {{``Digital Identity and Blockchain: A Comprehensive Overview of Approaches,"}},
    booktitle = {Research Handbook on Blockchain and Society},
    year = {2025. In Press},
    publisher = {De Gruyter}
}

@article{survey,
    author={Schumm, Daria and Müller, Katharina O. E. and Stiller, Burkhard},
    journal={IEEE Access}, 
    title={{Are We There Yet? A Study of Decentralized Identity Applications}}, 
    year={2025},
    volume={13},
    pages={125232-125259},
    doi = {https://doi.org/10.1109/access.2025.3588170}
}

@misc{a,
    title = {\textit{{``What is the Plan for Digital IDs and Will They be Mandatory?"}}},
    author = {Rachel Hagan},
    howpublished = {BBC. \url{https://www.bbc.com/news/articles/clyl3lzzed2o} (accessed on Dec. 18, 2025)}
}

@misc{b,
    title = {\textit{{``More than 1.6m Sign Petition Opposing Starmer’s Plan for Digital ID Cards."}}},
    author = {Kyriakos Petrakos},
    howpublished = {The Guardian. \url{https://www.theguardian.com/politics/2025/sep/27/petition-opposing-starmer-plan-digital-id-cards} (accessed Dec. 18, 2025)}
}

@misc{d,
    title = {\textit{{``Australia's Privacy Regulator Sues Optus over 2022 Data Breach."}}},
    howpublished = {Reuters. \url{https://www.reuters.com/business/media-telecom/australias-privacy-regulator-sues-optus-over-2022-data-breach-2025-08-08/} (accessed on Jan. 23, 2026)},
}

@article{2,
    title={Unleashing the power of internet of things and blockchain: A comprehensive analysis and future directions},
    author={Rejeb, Abderahman and Rejeb, Karim and Appolloni, Andrea and Jagtap, Sandeep and Iranmanesh, Mohammad and Alghamdi, Salem and Alhasawi, Yaser and Kayikci, Yasanur},
    journal={Internet of Things and Cyber-Physical Systems},
    volume={4},
    pages={1--18},
    year={2024},
    publisher={Elsevier}
}

@inproceedings{6,
    title={{Decentralized and Self-Sovereign Identity in the Era of Blockchain: A Survey}},
    author={Bai, Yirui and Lei, Hong and Li, Suozai and Gao, Haoyu and Li, Jun and Li, Leixiao},
    booktitle={IEEE International Conference on Blockchain},
    pages={500--507},
    year={2022},
    organization={IEEE},
    doi={https://doi.org/10.1109/blockchain55522.2022.00077}
}

@article{17,
    title={{Towards the Classification of Self-Sovereign Identity Properties}},
    author={{\v{C}}u{\v{c}}ko, {\v{S}}pela and Be{\'c}irovi{\'c}, {\v{S}}eila and Kami{\v{s}}ali{\'c}, Aida and Mrdovi{\'c}, Sa{\v{s}}a and Turkanovi{\'c}, Muhamed},
    journal={IEEE Access},
    volume={10},
    pages={88306--88329},
    year={2022},
    publisher={IEEE},
    doi = {https://doi.org/10.1109/access.2022.3199414}
}

@inproceedings{32,
    title={{Self-Sovereign identity ecosystems: benefits and challenges}},
    author={Laatikainen, Gabriella and Kolehmainen, Taija and Abrahamsson, Pekka},
    booktitle={12th Scandinavian Conference on Information Systems},
    year={2021},
    publisher={Association for Information Systems (AIS)},
    url={https://aisel.aisnet.org/scis2021/10}
}

@article{44,
    title={{Identity and Access Management Using Distributed Ledger Technology: A Survey}},
    author={Ghaffari, Fariba and Gilani, Komal and Bertin, Emmanuel and Crespi, Noel},
    journal={International Journal of Network Management},
    volume={32},
    number={2},
    pages={e2180},
    year={2022},
    publisher={Wiley Online Library},
    doi={https://doi.org/10.1002/nem.2180}
}

@article{51,
    title={{A Survey on Essential Components of a Self-Sovereign Identity}},
    author={M{\"u}hle, Alexander and Gr{\"u}ner, Andreas and Gayvoronskaya, Tatiana and Meinel, Christoph},
    journal={Computer Science Review},
    volume={30},
    pages={80--86},
    year={2018},
    publisher={Elsevier},
    doi={https://doi.org/10.1016/j.cosrev.2018.10.002}
}

@incollection{59,
    title={{Self-Sovereign Identity Systems: Evaluation Framework}},
    author={Satybaldy, Abylay and Nowostawski, Mariusz and Ellingsen, J{\o}rgen},
    booktitle={Privacy and Identity Management. Data for Better Living: AI and Privacy},
    editor={Michael Friedewald and Melek Önen and Eva Lievens and Stephan Krenn and Samuel Fricker},
    pages={447--461},
    year={2020},
    publisher={Springer}, 
    doi={https://doi.org/10.1007/978-3-030-42504-3_28}
}

@misc{61,
    author = {Allen, Christopher},
    title = {\textit{{``The Path to Self-Sovereign Identity."}}},
    howpublished = {\url{https://www.lifewithalacrity.com/article/the-path-to-self-soverereign-identity/} (accessed on May 12, 2024)}
}

@article{69,
    title={{A Systematic Literature Mapping on Using Blockchain Technology in Identity Management}},
    author={Ngo, TT Tram and Dang, T Anh and Huynh, V Vuong and Le, T Cong},
    journal={IEEE Access},
    volume={11},
    pages={26004--26032},
    year={2023},
    publisher={IEEE},
    doi={https://doi.org/10.1109/access.2023.3256519}
}

@article{70,
    title={{A First Look at Identity Management Schemes on the Blockchain}},
    author={Dunphy, Paul and Petitcolas, Fabien AP},
    journal={IEEE Security \& Privacy},
    volume={16},
    number={4},
    pages={20--29},
    year={2018},
    publisher={IEEE},
    doi = {https://doi.org/10.1109/msp.2018.3111247}
}

@inproceedings{72,
    title={{Decentralized Identity Systems: Architecture, Challenges, Solutions and Future Directions}},
    author={Dib, Omar and Toumi, Khalifa},
    booktitle={Annals of Emerging Technologies in Computing (AETiC), Print ISSN},
    pages={2516--0281},
    year={2020},
    doi={https://doi.org/10.33166/aetic.2020.05.002}
}

@article{79,
    title={{Blockchain-Based Identity Management System and Self-Sovereign Identity Ecosystem: A Comprehensive Survey}},
    author={Ahmed, Md Rayhan and Islam, AKM Muzahidul and Shatabda, Swakkhar and Islam, Salekul},
    journal={IEEE Access},
    volume={10},
    pages={113436--113481},
    year={2022},
    publisher={IEEE},
    doi={https://doi.org/10.1109/access.2022.3216643}
}

@article{84,
    title={{A Survey of Self-Sovereign Identity Ecosystem}},
    author={Soltani, Reza and Nguyen, Uyen Trang and An, Aijun},
    journal={Security and Communication Networks},
    volume={2021},
    number={1},
    pages={8873429},
    year={2021},
    publisher={Wiley Online Library},
    doi={https://doi.org/10.1155/2021/8873429}
}

@inproceedings{85,
    title={{Analysis of Identity Management Systems Using Blockchain Technology}},
    author={El Haddouti, Samia and El Kettani, M Dafir Ech-Cherif},
    booktitle={2019 International Conference on Advanced Communication Technologies and Networking (CommNet)},
    pages={1--7},
    year={2019},
    organization={IEEE},
    doi = {https://doi.org/10.1109/commnet.2019.8742375}
}

@article{87,
    title={{Blockchain-Based Identity Management: A Survey From the Enterprise and Ecosystem Perspective}},
    author={Kuperberg, Michael},
    journal={IEEE Transactions on Engineering Management},
    volume={67},
    number={4},
    pages={1008--1027},
    year={2019},
    publisher={IEEE},
    doi={https://doi.org/10.1109/tem.2019.2926471}
}

@article{89,
    title={{Decentralized and Self-Sovereign Identity: Systematic Mapping Study}},
    author={{\v{C}}u{\v{c}}ko, {\v{S}}pela and Turkanovi{\'c}, Muhamed},
    journal={IEEE Access},
    volume={9},
    pages={139009--139027},
    year={2021},
    publisher={IEEE},
    doi={https://doi.org/10.1109/access.2021.3117588}
}

@article{94,
    title={{In Search of Self-Sovereign Identity Leveraging Blockchain Technology}},
    author={Ferdous, Md Sadek and Chowdhury, Farida and Alassafi, Madini O},
    journal={IEEE Access},
    volume={7},
    pages={103059--103079},
    year={2019},
    publisher={IEEE},
    doi = {https://doi.org/10.1109/access.2019.2931173}
}

@article{102,
    title={{A Taxonomy of Challenges for Self-Sovereign Identity Systems}},
    author={Satybaldy, Abylay and Ferdous, Md Sadek and Nowostawski, Mariusz},
    journal={IEEE Access},
    year={2024},
    publisher={IEEE},
    doi={https://doi.org/10.1109/access.2024.3357940}
}

@inproceedings{186,
    title={{Blockchain-Based Self-Sovereign Identity: Survey, Requirements, Use-Cases, and Comparative Study}},
    author={Nokhbeh Zaeem, Razieh and Chang, Kai Chih and Huang, Teng-Chieh and Liau, David and Song, Wenting and Tyagi, Aditya and Khalil, Manah and Lamison, Michael and Pandey, Siddharth and Barber, K Suzanne},
    booktitle={IEEE/WIC/ACM International Conference on Web Intelligence and Intelligent Agent Technology},
    pages={128--135},
    year={2021},
    doi={https://doi.org/10.1145/3486622.3493917}
}

@article{189,
    title={{Blockchain-Based Identity Management Systems: A Review}},
    author={Liu, Yang and He, Debiao and Obaidat, Mohammad S and Kumar, Neeraj and Khan, Muhammad Khurram and Choo, Kim-Kwang Raymond},
    journal={Journal of Network and Computer Applications},
    volume={166},
    pages={102731},
    year={2020},
    publisher={Elsevier},
    doi={https://doi.org/10.1016/j.jnca.2020.102731}
}

@inproceedings{190,
    title={{A Comparative Survey on Blockchain-Based Self-Sovereign Identity System}},
    author={Kaneriya, Jayana and Patel, Hiren},
    booktitle={2020 3rd International Conference on Intelligent Sustainable Systems (ICISS)},
    pages={1150--1155},
    year={2020},
    organization={IEEE},
    doi={https://doi.org/10.1109/iciss49785.2020.9315899}
}

@article{191,
    title={{A Systematic Literature Mapping on Secure Identity Management Using Blockchain Technology}},
    author={Rathee, Tripti and Singh, Parvinder},
    journal={Journal of King Saud University - Computer and Information Sciences},
    volume={34},
    number={8},
    pages={5782--5796},
    year={2022},
    publisher={Elsevier},
    doi={https://doi.org/10.1016/j.jksuci.2021.03.005}
}

@article{192,
    author = {Tan, Kheng Leong and Chi, Chi-Hung and Lam, Kwok-Yan},
    title = {{Survey on Digital Sovereignty and Identity: From Digitization to Digitalization}},
    year = {2023},
    publisher = {Association for Computing Machinery},
    volume = {56},
    number = {3},
    journal = {ACM Computing Surveys},
    month = {oct},
    articleno = {61},
    numpages = {36},
    doi={https://doi.org/10.1145/3616400}
}

@incollection{193,
    title={{Decentralized Identity Management Using Blockchain Technology: Challenges and Solutions}},
    author={Buttar, Ahmed Mateen and Shahid, Muhammad Anwar and Arshad, Muhammad Nouman and Akbar, Muhammad Azeem},
    booktitle={Blockchain Transformations: Navigating the Decentralized Protocols Era},
    pages={131--166},
    year={2024},
    publisher={Springer},
    editor={Sheikh Mohammad Idrees and Mariusz Nowostawski},
    doi={https://doi.org/10.1007/978-3-031-49593-9_8}
}

@misc{198,
    title = {\textit{``The Laws of Identity,"}},
    author = {Cameron, Kim},
    howpublished = {\url {https://www.identityblog.com/stories/2005/05/13/TheLawsOfIdentity.pdf} (accessed on May 14, 2024)},
}

@misc{245,
    title={\textup{``Self-Sovereign Identity Solutions: The Necessity of Blockchain Technology,"}}, 
    author={Dirk van Bokkem and Rico Hageman and Gijs Koning and Luat Nguyen and Naqib Zarin},
    howpublished = {arXiv:1904.12816 [cs.CR], 2019. \url{https://arxiv.org/abs/1904.12816}} 
}

@inproceedings{249,
    title={{Establishing a Baseline for Evaluating Blockchain-Based Self-Sovereign Identity Systems: A Systematic Approach to Assess Capability, Compatibility and Interoperability}},
    author={Yao, Wei and Du, Wenlu and Gu, Jingyi and Ye, Junyi and Deek, Fadi P and Wang, Guiling},
    booktitle={Proceedings of the 2024 6th Blockchain and Internet of Things Conference},
    pages={108--119},
    year={2024},
    doi={https://doi.org/10.1145/3688225.3688239}
}

@misc{255,
    author = {Joe Tidy},
    title = {\textit{{``Ticketmaster Warns Customers to Take Action After Hack,"}}},
    howpublished = {BBC. \url{https://www.bbc.com/news/articles/c729e3qr48qo} (accessed on Dec. 18, 2025)}
}

@article{292,
    title={{Towards a Catalogue of Self-Sovereign Identity Design Patterns}},
    author={{\v{C}}u{\v{c}}ko, {\v{S}}pela and Ker{\v{s}}i{\v{c}}, Vid and Turkanovi{\'c}, Muhamed},
    journal={Applied Sciences},
    volume={13},
    number={9},
    pages={5395},
    year={2023},
    publisher={MDPI},
    doi={https://doi.org/10.3390/app13095395}
}

@inproceedings{294,
    title={{Design Patterns for Blockchain-Based Self-Sovereign Identity}},
    author={Liu, Yue and Lu, Qinghua and Paik, Hye-Young and Xu, Xiwei},
    booktitle={Proceedings of the European Conference on Pattern Languages of Programs 2020},
    pages={1--14},
    year={2020},
    doi={https://doi.org/10.1145/3424771.3424802}
}

@book{298,
    title={Requirements engineering fundamentals: a study guide for the certified professional for requirements engineering exam-foundation level-IREB compliant},
    author={Pohl, Klaus},
    year={2016},
    publisher={Rocky Nook, Inc.}
}

@inproceedings{318,
    title={{Modeling Security Requirements Through Ownership, Permission and Delegation}},
    author={Giorgini, Paolo and Massacci, Fabio and Mylopoulos, John and Zannone, Nicola},
    booktitle={13th IEEE International Conference on Requirements Engineering (RE'05)},
    pages={167--176},
    year={2005},
    organization={IEEE},
    doi={https://doi.org/10.1109/re.2005.43}
}

@incollection{319,
    title={{Filling the Gap Between Requirements Engineering and Public Key/Trust Management Infrastructures}},
    author={Giorgini, Paolo and Massacci, Fabio and Mylopoulos, John and Zannone, Nicola},
    booktitle={Public Key Infrastructure},
    pages={98--111},
    year={2004},
    editor={Sokratis K. Katsikas and Stefanos Gritzalis and Javier López},
    publisher={Springer}, 
    doi={https://doi.org/10.1007/978-3-540-25980-0_8}
}

@article{320,
    title={{Requirements Engineering Meets Trust Management: Model, Methodology, and Reasoning}},
    author={Giorgini, Paolo and Massacci, Fabio and Mylopoulos, John and Zannone, Nicola},
    journal={International Journal of Information Security},
    volume={5},
    pages={257–274},
    year={2006},
    organization={Springer},
    doi={https://doi.org/10.1007/s10207-006-0005-7}
}

@article{324,
    title={Predicate logic for software engineering},
    author={Parnas, David Lorge},
    journal={IEEE Transactions on Software Engineering},
    volume={19},
    number={9},
    pages={856--862},
    year={1993},
    publisher={IEEE}
}

@book{325,
    title={Non-functional requirements in software engineering},
    author={Chung, Lawrence and Nixon, Brian A and Yu, Eric and Mylopoulos, John},
    volume={5},
    year={2012},
    publisher={Springer Science \& Business Media}
}

@incollection{333,
    title={{A Generic Process for Requirements Engineering}},
    author={Hull, Elizabeth and Jackson, Ken and Dick, Jeremy},
    booktitle={Requirements Engineering},
    pages={23--46},
    year={2002},
    publisher={Springer},
    doi={https://doi.org/10.1007/1-84628-075-3_2}
}

@incollection{334,
    title={{Writing and Reviewing Requirements}},
    author={Hull, Elizabeth and Jackson, Ken and Dick, Jeremy},
    booktitle={{Requirements Engineering}},
    pages={93-111},
    year={2017},
    publisher={Springer},
    doi={https://doi.org/10.1007/978-3-319-61073-3_4}
}

@misc{337,
    author={{Intersoft Consulting}},
    title={GDPR Art. 6 Lawfulness of Processing},
    year={n.d.}, 
    url={https://gdpr-info.eu/art-6-gdpr/},
    accessed={24 July 2025},
}

@misc{338,
    author={{Intersoft Consulting}},
    title={GDPR Art. 7 Conditions for Consent},
    year={n.d.}, 
    url={https://gdpr-info.eu/art-7-gdpr/},
    accessed={24 July 2025},
}

@misc{339,
    author={{Intersoft Consulting}},
    title={Recital 32: Conditions for Consent},
    year={n.d.}, 
    url={https://gdpr-info.eu/recitals/no-32/},
    accessed={24 July 2025},
}






\end{document}